\documentclass[pra,twocolumn,showpacs,preprintnumbers,floatfix]{revtex4}
\usepackage{amsfonts}
\usepackage{amsmath}
\usepackage{amssymb}
\usepackage{graphicx}
\usepackage{dcolumn}
\usepackage{bm}%
\begin{document}
\title[Short title for running header]{Excitations of Bose-Einstein condensates in a one-dimensional
periodic potential}
\author{N.~Fabbri*, D.~Cl\'ement, L.~Fallani, C.~Fort, M.~Modugno, K.~M.~R.~van~der~Stam, M.~Inguscio}
\affiliation{European laboratory for Non-linear Spectroscopy (LENS), Dipartimento di Fisica - Universit\`a
di Firenze and INFM-CNR, via N. Carrara 1, 50019 Sesto Fiorentino (FI), Italy}


\begin{abstract}
We report on the experimental investigation of the response of a three-dimensional Bose-Einstein condensate (BEC) in the presence of a one-dimensional (1D) optical lattice. By means of Bragg spectroscopy we probe the band structure of the excitation spectrum in the presence of the periodic potential. We selectively induce elementary excitations of the BEC choosing the transferred momentum and we observe different resonances in the energy transfer, corresponding to the transitions
to different bands. The frequency, the width and the strength of these resonances are investigated as a function of the amplitude of the 1D optical lattice.
\end{abstract}

\maketitle

The knowledge of the linear response of a complex system gives crucial information about its many-body behavior.
For example, the superfluid properties of a three-dimensional (3D) Bose-Einstein condensate (BEC) are related to the linear part of
the phononic dispersion relation at low momenta \cite{Stringari}. The presence of optical lattices enriches
the excitation spectrum of a BEC in a remarkable way. For deep three-dimensional lattices, the gas enters the strongly
correlated Mott insulator phase and the spectrum exhibits a gap at low energies \cite{Mottspectrum}.
The response of a BEC in the superfluid phase is also drastically modified by the presence of a one-dimensional (1D) optical lattice \cite{Sorensen1998,Wu2002,Menotti2003,Kraemer2003,Modugno2004}. Indeed, as in any periodic system, energy gaps open in the spectrum at the multiples of the lattice momentum and it is possible to excite several states corresponding to different energy bands at a given value of the momentum transfer \cite{Denschlag2002,Eiermann}. In addition, the linear dispersion relation of the superfluid, and thus its sound velocity, is changed. In the mean-field regime of interactions these peculiar features of the excitations of a superfluid BEC in the presence of an optical lattice are captured by the Bogoliubov theory \cite{Stringari}.

\begin{figure}[ht!]
\includegraphics[width=0.68\columnwidth]{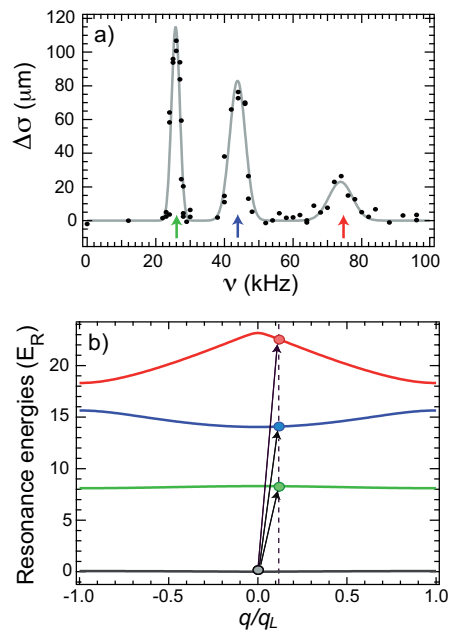} \caption{(a) Measured BEC excitation spectrum in the presence of a lattice with height $s=(22\pm2)$.
The increase $\Delta\sigma$ of the width of the atomic density distribution is monitored as a function of
the relative detuning $\nu$ between the two
counterpropagating Bragg beams. The data are fitted with Gaussian functions (black line). The arrows below the resonances
indicate the corresponding bands, represented in (b) with the same colors.
(b) Band structure of the excitation spectrum of a BEC in a 1D optical lattice with $s=22$: first, second, third
and fourth bands are represented (gray, green, blue and red lines respectively). The arrows indicate
the processes starting from a BEC at $q=0$ and inducing the creation of excitations in the
different bands at a quasi-momentum transfer $0.12q_L$.} \label{Fig1}
\end{figure}

Bragg spectroscopy represents an excellent experimental tool to investigate the linear
response of gaseous BECs \cite{Davidson}. It has allowed to measure the dispersion relation of interacting BECs in the mean-field regime \cite{Stenger1999,Steinhauer2002,Steinhauer2003}, to characterize the presence of phase fluctuations in elongated BECs \cite{Richard2003}, to study signatures of vortices \cite{Muniz2006} and more recently to study strongly interacting 3D Bose \cite{Papp2008} and Fermi \cite{Veeravalli2008} gases close to Feshbach resonances as well as 1D Bose gases across the superfluid to Mott insulator transition \cite{noiSFMI}.

In this work we use Bragg spectroscopy to probe the excitation spectrum of a 3D BEC loaded in a 1D optical lattice.
Previous experimental studies have so far investigated the excitations of superfluid BECs within the lowest energy band of a 3D optical lattice by means of lattice modulation \cite{Schori2004}
and Bragg spectroscopy \cite{noiSFMI,Du2007}. This paper presents a detailed experimental study of the different bands in the excitation
spectrum of an interacting 3D BEC in the presence of a 1D optical lattice.
We measure the resonance frequencies, the strengths and the widths
of the transitions to different bands of the 1D optical lattice. The measurements are quantitatively compared with Bogoliubov mean-field calculations for our experimental system \cite{Modugno2004}.

We produce a 3D cigar-shaped BEC of N$\simeq3\times10^{5}$ $^{87}$Rb atoms in a Ioffe-Pritchard
magnetic trap whose axial and radial frequencies are $\omega_{y}=2\pi
\times8.9$ Hz and $\omega_{x}=\omega_{z}=2\pi\times90$~Hz respectively,
corresponding to a chemical potential $\mu\simeq h\times1$~kHz, with $h$ being the Planck constant.
The condensate is loaded in an optical lattice along the longitudinal direction ($\hat{y}$ axis). Two counterpropagating laser beams with wavelength $\lambda_{\mathrm{L}}=830$ nm create the lattice potential
$V(y)=s E_{R} \sin^{2}(q_{\mathrm{L}}y)$ where
$q_{\mathrm{L}}=2 \pi /\lambda_{\mathrm{L}}=7.57~\mu$m$^{-1}$ is
the wave-number of the beams, $s$ measures the height of
the lattice in units of the recoil energy $E_{R}=h^{2}/2m\lambda_{\mathrm{L}}^{2}\simeq
h\times3.3$~kHz, $m$ is the mass of a $^{87}$Rb atom.
The loading of the BEC in the lattice is performed by slowly
increasing the laser intensity up to a height $s$ with a $140$ ms long exponential ramp with time constant $\tau=30$ ms.

After a holding time of typically $20$ ms in the lattice at the height $s$, we excite the gas by shining two off-resonant laser beams (Bragg beams) for a time $\Delta t_{B}=3$ ms.
The Bragg beams induce a two-photon transition transferring momentum and energy to the atomic sample. Their wavelength is $\lambda_{B}=780$ nm corresponding to a wave-number $q_{B}=2\pi/\lambda_{B}=8.05~\mu$m$^{-1}$, and they are typically
detuned by $350$ GHz with respect to the D$2$ transition of $^{87}$Rb.
To change the transferred momentum we use two different geometries of the Bragg beams. In the first configuration the two
beams are counter-propagating along the $\hat{y}$ direction and the transferred momentum is $q=2q_B=2.12q_L$,
that corresponds to a quasi-momentum $0.12q_L$. In the second configuration the angle between the Bragg beams is smaller and the measured value of the transferred momentum (and quasi-momentum, in this case) along the $\hat{y}$ direction is $q=0.96q_L$.
In both the cases, the two beams are detuned from each other by a frequency difference $\nu$ using two phase-locked acousto-optic modulators. We quantify the response to the excitation by measuring the energy transferred to the gaseous BEC. The measurement of the energy transfer $E(\nu, q)$ is connected with the dynamical structure factor $S(\nu, q)$ (giving information on the excitation spectrum) by the relation \cite{Stringari}
\begin{equation}
E(\nu, q) \propto \nu S(\nu, q), \label{Eq1}%
\end{equation}
where $\nu$ and $q$ are the frequency and the momentum of the excitation. In particular, this result applies for long enough Bragg pulses, namely $\nu \Delta t_B \gg 1$, which is the case in our experiment since $\nu$ is of the order of several kHz.

In order to get an estimate of the transferred energy $E(\nu, q)$, we adopt the following procedure. We linearly ramp in $15$~ms the longitudinal optical lattice from $s$ (the lattice height at which we have applied the Bragg pulse) to the fixed value $s_{f}=5$.
Then we let the excitation being redistributed over the entire system by means of the inter-atomic collisions for $5$~ms. After this interval time we abruptly switch off both the optical
lattice and the magnetic trap, letting the cloud expand for a time-of-flight $t_{\mathrm{TOF}}=20$ ms
and we then take an absorption image of the density distribution integrated along the $\hat{x}$ axis. Since the atoms are released from an optical lattice of relatively small amplitude ($s_{f}=5$), the density distribution exhibits an interference pattern \cite{greiner2001}. We extract the \emph{rms} width $\sigma$ of the central peak of this density distribution
by fitting it with a Gaussian function. The increase $\Delta\sigma$ of this quantity is used as a measurement of the energy transfer \cite{Schori2004}.
For a given value of the transferred momentum $q$ and amplitude $s$ of the lattice, this procedure is repeated varying the energy $h \nu$ of the excitation in order to
obtain the spectrum. In our regime of weak inter-atomic interactions, the excitation spectrum of the BEC in the presence of a 1D optical lattice can be described by the
mean-field Bogoliubov approach \cite{Sorensen1998,Menotti2003}, by which we calculate the resonance frequencies $\nu_{\mathrm{j}}$
and the transition strengths $Z_{\mathrm{j}}$ to create an excitation in the Bogoliubov band $j$.

\begin{figure}[ht!]
\includegraphics[width=0.64\columnwidth]{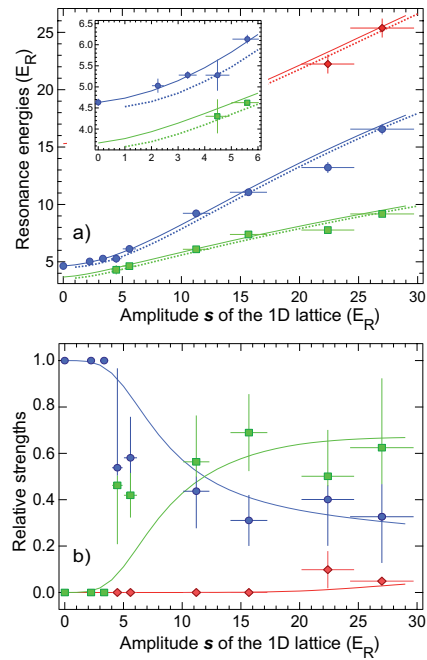} \caption{(a)
Band spectroscopy of a BEC in the presence of a 1D optical lattice: the energy of the resonances is reported
as a function of the height $s$ of the lattice. The experimental points (green squares,
blue circles and red diamonds) are compared with the numerical calculation of the Bogoliubov spectrum
in the presence of a 1D lattice (solid lines) and the single-particle Bloch spectrum (dashed lines). The lines correspond to the energy of an excitation in the second (green line), third (blue line)
and fourth (red line) Bogoliubov bands. In the inset of (a): zoom of the graph (a) for low values of $s$. (b)
Relative strength of the excitations in the $\mathrm{j}=2$, $\mathrm{j}=3$ and $\mathrm{j}=4$ bands. Symbols and colors are the same as in (a).} \label{Fig2}
\end{figure}

We first discuss the results obtained with the configuration of counter-propagating beams, i.e. for a transferred momentum $q=2.12q_L$. The induced two-photon transition is characterized by a measured Rabi frequency for the BEC in the absence of the optical lattice $\Omega_R\simeq2\pi\times1$ kHz for the typical power and detuning of the beams used in the experiment. A typical Bragg spectrum is presented in Fig.~\ref{Fig1}(a) corresponding to a lattice height $s=(22\pm2)$. The spectrum exhibits
multiple resonances corresponding to the creation of excitations in the different Bogoliubov bands as shown in Fig.~\ref{Fig1}(b). From Gaussian fit of each resonance we extract the central frequency, the width and the relative strength of the transition towards the corresponding band. In Fig.~\ref{Fig2}(a), we plot the energy values corresponding to the measured central frequencies as a function of $s$. The vertical error bars come from the result of the fitting procedure while the horizontal error bars correspond to possible systematic errors in the lattice calibration (estimated within 10\%).
For large enough amplitude $s$ of the periodic potential we observe up to three different bands.
We find a good agreement between the experimental data and the numerical results of the Bogoliubov calculation (solid lines in Fig.~\ref{Fig2}(a)). In particular, for low amplitudes of the 1D lattice ($s<6$)
the agreement of the resonance energies with the Bogoliubov bands (full lines) is
better than with the single-particle (dashed lines) Bloch bands (see inset in Fig.~\ref{Fig2}(a)). For larger amplitude of the 1D lattice, we can not explicitly distinguish between the Bogoliubov and Bloch results. This comes from the experimental uncertainty on the calibration of the lattice amplitude.

In the entire range of $s$ values used in this work, we observe a resonance corresponding to an excitation created in the third band $\mathrm{j}=3$ (blue circles on Fig.~\ref{Fig2}).
For larger lattice amplitudes two other resonances appear, respectively for $s>4$ and $s>20$, corresponding
to an excitation in the $\mathrm{j}=2$ band (green squared in Fig.~\ref{Fig2}(a))
and in the $\mathrm{j}=4$ band (red diamonds in Fig.~\ref{Fig2}(a)). This demonstrates the possibility to excite, in a
periodic system, several states for a given momentum transfer \cite{Denschlag2002}.
For weak optical lattices creating an excitation in the $\mathrm{j}=3$ band is the most efficient process since the excitation energy of this band is continuously connected as $s\rightarrow0$ to that of the BEC in the absence of the 1D
optical lattice at the transferred momentum $q=2.12q_L$. On the contrary, the possibility to excite states in the second and fourth bands of the optical lattice requires a large enough amplitude $s$. These observations can be quantified in terms of the strength $Z_{\mathrm{j}}$ of the different excitations, which can be extracted from the energy spectrum. The strengths $Z_{\mathrm{j}}$ are proportional to the integral $\int d\nu \ S_{\mathrm{j}}(q,\nu)$ with $S_{\mathrm{j}}(q,\nu)$ being the structure factor corresponding to the creation of an excitation in the Bogoliubov band $\mathrm{j}$ \cite{Menotti2003}.
From Eq.~\ref{Eq1} and assuming that $\nu_{\mathrm{j}}$ is much larger than the width of the resonances of $S_{\mathrm{j}}(q,\nu)$,we obtain

\begin{equation}
Z_{\mathrm{j}}(q)\propto \int d\nu \ S_{\mathrm{j}}(q,\nu)\propto \frac{1}{\nu_{\mathrm{j}}}\int d\nu \ E_{\mathrm{j}}(q,\nu)\equiv g_{\mathrm{j}}.
\label{Eq2}
\end{equation}

In the experiment, we extract the quantity $g_{j}$ from a Gaussian fit of the different resonances. Normalizing the sum of these quantities to one for the first three observed resonances $(g_{2}+g_{3}+g_{4}=1)$ allows direct comparison
with the relative strengths $Z_{j}/(Z_{2}+Z_{3}+Z_{4})$ for $j=2,3,4$. The comparison between the experimental data
and the calculation reveals a reasonable agreement (see Fig.~\ref{Fig2}(b)).

\begin{figure}[ht!]
\includegraphics[width=0.7\columnwidth]{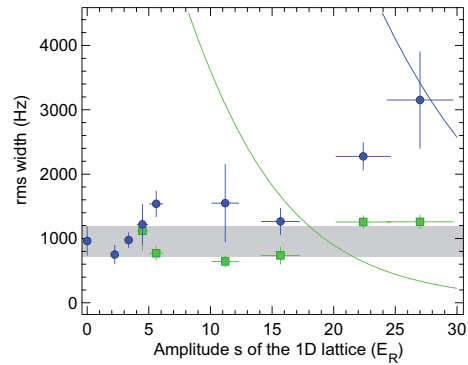}
\caption{\emph{Rms} width of the resonances $\mathrm{j}=2$ (green squared) and
$\mathrm{j}=3$ (blue circles) as a function of $s$. The
gray region corresponds to the experimental \emph{rms} width (with its uncertainty)
for the BEC in the absence of the lattice ($s=0$). The green
and blue lines are respectively the bandwidth of the second and the third band,
calculated in the mean-field Bogoliubov approach.} \label{Fig3}
\end{figure}

From the Gaussian fit of the experimental spectra (see Fig.~\ref{Fig1}(c)), we also extract the \textit{rms} width of the resonances $\mathrm{j}=2$ and $\mathrm{j}=3$ with the results plotted in Fig.~\ref{Fig3}.
Different sources contribute to broaden the observed resonances. The inhomogeneous density of the trapped BEC is a first source \cite{Stenger1999}. From the measured spectrum of the BEC in the absence of the optical lattice ($s=0$) we extract this contribution being $(0.36 \pm 0.11)$ kHz, consistent with the expected value $\simeq0.26$ kHz \cite{Stenger1999}. The other contributions to the width are related to the Bragg spectroscopic scheme. For our experimental parameters the largest contribution comes from the power broadening ($\Delta\nu_P\simeq1$ kHz), whereas the atom-light interaction time broadening ($\Delta\nu_t\simeq167$ Hz) is much smaller. The total resonance width can be obtained by quadratically adding up all these \emph{rms} contributions.
In the presence of the optical lattice we observe that the widths of the resonances corresponding to the excitations in the bands $\mathrm{j}=2$ and $\mathrm{j}=3$ lie within the experimental range of the resolution as expected for a coherent system, except in the case of $\mathrm{j}=3$ for large amplitudes of the lattice ($s>20$) where the width is much larger.
We attribute these larger widths at high amplitude of the 1D lattice
to the long tunneling times ($0.11$~s for $s=20$) implying that the system is not fully coherent along the $\hat{y}$ direction on the time scale of the experiment. Indeed, the loss of coherence spreads the population of quasi-momenta across a larger fraction of the Brillouin zone. This results in a wider range of resonant energies in the system and, for large amplitude $s$, one expects the width of the resonances in the energy spectrum to be equal to the bandwidths. In Fig.~\ref{Fig3} we have plotted the bandwidths of the $\mathrm{j}=2$ and $\mathrm{j}=3$ bands (green and blue lines). When the system becomes incoherent the width of the resonance $\mathrm{j}=3$ is equal to the bandwidth.
This effect is not observable for the band $\mathrm{j}=2$ where the bandwidth (green line in Fig.~\ref{Fig3}) is smaller than the experimental resolution.

\begin{figure}[ht!]
\includegraphics[width=0.68\columnwidth]{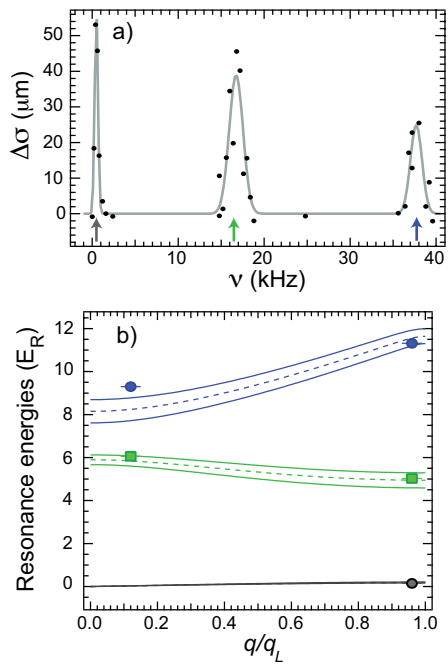}
\caption{(a) Excitation spectrum of a BEC in the presence of a 1D lattice with height $s=11$ at a
transferred momentum $q=0.96q_L$ along the $\hat{y}$ direction. The arrows below the resonances
indicate the corresponding bands, represented in (b) with the same colors. (b) Energy of the resonances corresponding to an excitation in the band $\mathrm{j}=1$ (gray circle), $\mathrm{j}=2$ (green squared) and
$\mathrm{j}=3$ (blue circle) as a function of the transferred quasi-momentum $q$ for a fixed value of the lattice height $s=(11\pm1)$. The experimental points are compared with the numerical calculation of the energy bands in the Bogoliubov approach for $s=11$ (gray, green and blue dotted lines); the solid lines correspond
to the bands for $s=10$ and $s=12$ to take into account the $10\%$ uncertainty of $s$.} \label{Fig4}
\end{figure}

We also perform the experiment with a different configuration of the Bragg beams corresponding to a transferred momentum along $\hat{y}$ $q=0.96q_L$. In Fig.~\ref{Fig4}(a) an excitation spectrum in the presence of an optical lattice of height $s=11$ is depicted. Note that a first resonance at low frequency is visible corresponding to an excitation with non-zero momentum within the lowest energy band ($\mathrm{j}=1$). Such a resonance is not observed using counter-propagating Bragg beams because the strength of this transition is negligible for $q=2.12q_L$. Due to the variation of the transferred momentum $q$ with respect to the previous case, the frequency of the resonances are shifted according to the dispersion relation of the different energy bands of the system. In Fig.~\ref{Fig4}(b) we report the frequency of the resonances $\mathrm{j}=1$, $\mathrm{j}=2$ and $\mathrm{j}=3$ for the two values of quasi-momentum used in the experiment ($0.12 q_L$ and $0.96 q_L$). We use the region comprised between the calculated bands for $s=10$ and $s=12$ (solid lines) to take into account the $10\%$ error in the lattice calibration. The experimental points are in good agreement with the numerical calculation of the Bogoliubov bands for $s=11$ (dotted lines in the Figure) within the experimental uncertainty.

In conclusion, Bragg spectroscopy has been used to probe the response of a Bose-Einstein condensate in the presence of a 1D optical lattice. Changing the angle of the Bragg beams allowed us to investigate excitations for a transferred quasi-momentum close to the center and to the edge of the reduced Brillouin zone. We have observed different resonances in the response function of the system corresponding to the different bands of the periodic potential. Being the system in a weakly interacting regime, the experimental results are in quantitative agreement with the Bogoliubov bands approach. This work opens the way to investigate the modification of the excitation spectrum in the presence of an additional lattice with different wavelength (bichromatic potential) \cite{diener2001} and eventually to study the localization of the excitations in a true disordered potential \cite{Lugan}.

This work has been supported by UE contract No. RII3-CT-2003-506350, MIUR PRIN 2007 and Ente Cassa di Risparmio di Firenze, DQS EuroQUAM Project, NAMEQUAM Project and Integrated Project SCALA. We acknowledge all the colleagues of the Quantum Degenerate Group at LENS for fruitful comments.

\end{document}